\documentclass[aps,superscriptaddress,showpacs,preprintnumbers,twocolumn]{revtex4}
\usepackage{amsmath}
\usepackage{graphicx}
\usepackage{draftcopy}
\begin{document}
\preprint{LA-UR-04-6667}
\title{Inverse-Scattering Theory and the Density Perturbations from Inflation}

\author{Salman Habib}
\affiliation{T-8, University of California, Los Alamos National 
Laboratory, Los Alamos, New Mexico 87545, USA}

\author{Katrin Heitmann}
\affiliation{ISR-1, University of California, Los Alamos National 
Laboratory, Los Alamos, New Mexico 87545, USA}

\author{Gerard Jungman}
\affiliation{T-6, University of California, Los Alamos National 
Laboratory, Los Alamos, New Mexico 87545, USA}

\date{\today}

\begin{abstract}

Inflationary cosmology provides a successful paradigm for solving
several problems, notably the generation of density perturbations
which seed the formation of observed cosmic structure.  We show how to
use inverse scattering theory as the basis for the inflationary
reconstruction program, the goal of which is to gain information about
the physics which drives inflation.  Inverse scattering theory
provides an effective and well-motivated procedure, having a sound
mathematical basis and being of sufficient generality that it can be
considered the foundation for a non-parametric reconstruction
program. We show how simple properties of the power spectrum translate
directly into statements about the evolution of the background
geometry during inflation.

\end{abstract}

\pacs{98.80.Cq}

\maketitle


A wealth of data drives much of the current activity in cosmology.
Amongst these data, perhaps the cleanest signal from the primordial
Universe is the power spectrum of density fluctuations, which is
obtained from observations of fluctuations in the microwave background
sky \cite{cmbobs} and from observations of the large-scale clustering
of matter \cite{lssobs}.  Especially in the case of microwave
background observations, the observed fluctuations give direct
information on the ``initial conditions'' for the density
perturbations which manifest in the gravitational clustering of matter
in the Universe.

In an inflationary universe, these initial density perturbations arise
as a relic of quantum fluctuations from the very earliest times.  The
inflationary evolution causes these modes to grow in amplitude, with a
time-evolution that is determined completely by the evolution of the
space-time geometry, encoded in the scale-factor \(a(t)\). This
evolution imprints a signature of the inflationary evolution on the
spectrum of modes.  Therefore, structure in the mode spectrum can be
directly related to the evolution of \(a(t)\) and hence to the nature
of the stress-energy which drives inflation.  Because the nature of
inflation is quite mysterious, information of this sort is very
important for cosmology.

Consider now the standard picture for quantum fluctuations
and the generation of structure during inflation. The equation
describing the evolution of gauge-invariant metric fluctuations
can be taken in the form \cite{Mukhanovetal}
\begin{equation}
  \phi_k'' + \left[k^2 - q(\eta) \right] \phi_k = 0,
\label{eqn:basic}
\end{equation}
where primes indicate derivatives with respect to conformal time
\(\eta\), and \(q(\eta)\) is an effective potential.  The spatial
dependence of the functions has been expanded in modes specified by
three-momentum \(k\), so that \(\phi_k\) is a function of \(\eta\)
only.  For scalar modes we have \(q(\eta) = z''/z\), where \(z(\eta)\)
is a function determined by the evolving background; under quite
general circumstances it is given by
\[
  z(\eta) = \frac{a(\eta)}{c_s(\eta) h(\eta)}
            \left[h^2(\eta) - h'(\eta) + \mathcal{K}\right]^{1/2},
\]
where \(h(\eta) \equiv a'(\eta)/a(\eta)\) and \(c_s(\eta)\) is the
``adiabatic sound speed''.  The curvature term \(\mathcal{K}\) rapidly
becomes negligible as inflation proceeds and will be ignored for the
remainder of this Letter.  When inflation is driven by a scalar field,
\(c_s = 1\).  For tensor modes the effective potential is given by
\(q_T(\eta) = a''(\eta)/a(\eta)\).  In the infinite past, the mode
population is that of a precisely specified vacuum state, which gives
the condition
\[
  \phi_k \sim (2k)^{-\frac{1}{2}}\; e^{-i k \eta + i \gamma},
  \quad \eta\rightarrow -\infty.
\]
The argument of the exponential is determined by the positive-energy
condition in the deep past, and the normalization is determined by the
usual free-field normalization. The constant phase factor \(\gamma\)
is a matter of convention and is fixed later in order to simplify
resulting expressions.

A given \(k\)-mode evolves to a good approximation as a free mode
until it begins to feel the effective potential, for \(\eta\) such
that \(k^2 \simeq q(\eta)\).  When it crosses into the forbidden
region (crosses outside the Hubble length), it begins to grow in
amplitude. Deep within the forbidden region, it can be easily shown
that the scalar modes behave like \(\phi_k \sim A_k z(\eta)\), where
\(A_k\) is a function of \(k\) only.  The structure of \(q(\eta)\) is
imprinted on this function, which is directly related to the power
spectrum of the fluctuations \(\phi_k\),
\[
  P(k) = \frac{k^3}{2\pi^2} \left|A_k \right|^2.
\]
This power spectrum is in turn directly related to the spectrum of
density fluctuations or curvature fluctuations.

The goal of inflationary reconstruction is to determine as much
information as possible about the stress-energy which drove the
inflationary expansion. The reconstruction program splits naturally
into two clearly defined tasks.  The first task is to solve the linear
problem, inverting from the observed power spectrum to the function
\(q(\eta)\). The second task is to use the obtained \(q(\eta)\) to
constrain the geometry and therefore to constrain the stress-energy
tensor which drives inflation. In this Letter we show how to use
inverse scattering theory to complete the first task; we then show how
this information can be used in the second task, to constrain the
physics of the inflationary epoch.


The basic problem of calculating density perturbations is most
naturally posed as a problem in scattering theory for the wave
equation (\ref{eqn:basic}).  In the following we can restrict
ourselves to the case of an effective radial Schr\"odinger equation
for scattering from a central potential. In partial wave \(\ell\) this
equation is
\[
  \phi''(k,r) + \left[k^2 - \frac{\ell(\ell+1)}{r^2} - V(r)\right] \phi(k,r) = 0.
\]
In order to avoid confusion in applying the formalism to the wave
equations given above, we identify \(r = -\eta\) and work solely in
terms of the variable \(r\).

The reason to concentrate on the partial wave equation at fixed
\(\ell\) is the following. It turns out to be most natural to split
the potential in the form
\[
  q(r) = \frac{\nu^2(r) - \frac{1}{4}}{r^2}
       = \frac{\bar\nu^2 - \frac{1}{4}}{r^2} + \frac{\nu^2(r) - \bar\nu^2}{r^2},
\]
where \(\bar\nu\) is a judiciously chosen constant.  We identify
\(\ell(\ell+1) = \bar\nu^2 - 1/4\) and the remaining term as the
central potential \(V(r)\). In this way we collect the most singular
part of the potential into the effective ``centrifugal'' term. The
remainder of the discussion explains precisely how this process is
carried out and what it means.  Note also that there are no bound
states in the problem, independent of the sign of \(V(r)\), because of
the positivity of \(q(\eta)\).

The wave solution is purely an incoming wave for
\(r\rightarrow\infty\), which follows from the positive-energy vacuum
condition in the deep past.  Solutions which obey this condition are
important in scattering theory and are called Jost solutions.
Precisely, a Jost solution is the unique solution of the wave equation
satisfying the condition \(\phi(k,r) \sim \exp(ikr + i\pi\ell/2)\),
\(r\rightarrow\infty\).  Note that a Jost solution must become
singular deep inside the forbidden region, as a consequence of the
lack of reflection in its definition. In our case the Jost-like
solution is actually the physically interesting solution, describing a
positive-energy quantum mode from the deep past.

The behaviour of a Jost solution deep inside the forbidden region is
described by a function of \(k\),
\[
  F_\ell(k) = \lim_{r\rightarrow0}
    \left[
    \frac{e^{-i\pi\ell}\Gamma(\frac{1}{2})}{\Gamma(\ell+\frac{1}{2})}
    \left(\frac{kr}{2}\right)^\ell f_\ell(k, r) \right], 
\]
where \(f_\ell(k, r)\) is the Jost solution for \(k\).  The function
\(F_\ell(k)\) is the so-called Jost function, and it encodes
essentially all the information about the scattering theory.  Given
our setup of the problem, we have \(z \sim c_0 (r_0/r)^{\ell}\) for
\(r\rightarrow 0\), where \(c_0\) is a \(k\)-independent constant.
Therefore a direct relation holds between the Jost function
\(F_\ell(k)\) and the basic function of interest \(A_k\),
\[
  A_k = \frac{\Gamma(\ell+\frac{1}{2})}{c_0 \Gamma(\frac{1}{2})} \;
        \left(\frac{2}{k}\right)^{\ell+1/2} F_\ell(k).
\]

At this point it is possible to invoke the full machinery of inverse
scattering theory, which allows us to reconstruct the potential
\(V(r)\) from the Jost function, by solving a linear integral equation
\cite{ChadanSabatier}. Rather than solve this numerical problem, in
this Letter we instead show how the Jost function encodes information
about the asymptotic behaviour of \(V(r)\), and therefore the function
\(z(r)\), in a manner which is directly applicable to the cosmological
inversion problem.

To be precise, suppose that we write \(z(r) = (r_0/r)^\ell c(r)\),
where \(c(r)\) is assumed to be regular and to satisfy \(c(\infty) =
1\).  The potential is then given by
\[
  V(r) = \frac{z''}{z} - \frac{\ell(\ell+1)}{r^2}
       = \frac{c''}{c} - \frac{2\ell}{r} \frac{c'}{c}.
\]
By direct calculation, we can show that \(F_\ell(0) = c(0)\).  This is
a consequence of the special form for \(V(r)\).  A more difficult
result expresses the behaviour of \(F_\ell(k)\) for \(k\rightarrow 0\)
in terms of the large \(r\) behaviour of \(V(r)\). To this end we
define the length-scale \(r_1\) and the exponent \(\delta\) by
\[
  V(r) \sim \frac{1}{r^2} \left(\frac{r_1}{r}\right)^\delta,\quad
  r\rightarrow\infty. 
\]
The small \(k\) behaviour for \(\ell=0\) has been calculated by Klaus
\cite{Klaus}. Extending Klaus' calculation to general \(\ell\), we
obtain the \(k\rightarrow 0\) asymptotic
\begin{equation}
  \frac{F_\ell(k)}{F_\ell(0)} = 1 +
    e^{-i\pi\delta/2}
    \frac{\Gamma(\frac{1}{2}) A(\ell,\delta)}{\Gamma(\ell +
    \frac{3}{2}) 2^{\ell+1}} 
    (k r_1)^\delta + \cdots,
\label{eqn:KlausEXT}
\end{equation}
where \(A(\ell,\delta) = A_1(\ell,\delta) + A_2(\ell,\delta)\) with
\begin{eqnarray*}
  A_1(\ell, \delta)
    &=& 2^{\ell-1-\delta}
        \frac{\Gamma(\ell+\frac{1}{2} -
    \frac{\delta}{2})}{\Gamma(\frac{1}{2})}, \\ 
  A_2(\ell, \delta)
    &=& \Gamma(\ell-\delta)
    \left[2^{\ell+1+\delta}
          \frac{\Gamma(\ell+\frac{3}{2})
    \Gamma(1+\delta)}{\Gamma(\frac{1}{2}) \Gamma(2+\ell+\delta)} 
          - 1
    \right].
\end{eqnarray*}
This particular calculation assumes \(0 < \delta < 1\), which turns
out to be the generic case for inflationary models; other cases can be
handled in a similar manner.  Given this result, we see that knowledge
of the Jost function for \(k\) near zero translates into knowledge of
\(V(r)\) and \(z(r)\) for large \(r\).  This result is all we will
need for the remainder of the discussion.


Before proceeding let us examine the meaning of the various
ingredients introduced thus far. First consider the form chosen for
the potential. In power-law inflation, where the scale-factor is
assumed to grow like \(a(t) \sim t^p\), the function \(\nu(r)\) is a
constant, \(\nu = 3/2 + 1/(p-1)\).  The fact that the power-law case
corresponds precisely to a free wave equation is essentially a
statement of the exact scale-invariance of the wave equation in that
case.  The usual spectral index and the partial wave index are related
by \(n = 3 - 2\ell.\) The Harrison-Zeldovich case with \(n=1\)
corresponds to partial wave \(\ell=1\).  When \(\ell\) is different
from zero, the power-spectrum has a pure power-law scaling form with
an anomalous scaling exponent.  It is interesting to note that the
Harrison-Zeldovich case corresponds to an anomalous exponent, although
from another point of view it is ``natural''.

Only deviations from scale-invariance in the wave equation can give
rise to deviations from power-law behaviour in the spectrum of density
perturbations and to a nontrivial Jost function. Clearly the Jost
function is identically one in the free case. Note that the parameter
\(\ell\) need not be an integer; nothing in the analysis is affected
by this.

Next consider the meaning of the limit \(r\rightarrow 0\) which
defines the Jost function and the spectrum of density
perturbations. Obviously the detailed structure of the potential in a
neighbourhood of the origin depends on the way in which inflation
ends; if the details of this process become important, then we lose
the ability to make robust predictions about the spectrum of density
perturbations. However, this is simply a statement defining the range
of \(k\) modes of interest. In practice we observe a range of modes,
which we assume were ``well-inflated'', so that they are not sensitive
to such details. A typical complementarity relation holds, so that
this range of \(k\) modes corresponds to a range of \(r\). In this
spirit, the range of \(k\) modes of interest corresponds to {\em
small} \(k\) and therefore to large \(r\); these are the modes which
spent a significant amount of time outside the horizon.

Combining the expressions given above, we obtain a representation for
the power spectrum, 
\begin{equation}
  P(k) = \frac{|\Gamma(\ell+\frac{1}{2})|^2}{4 \pi^3}
         \; k^2 \left(\frac{k r_0}{2}\right)^{-2\ell}
         \left|\frac{F_\ell(k)}{F_\ell(0)}\right|^2. 
\label{eqn:Pk}
\end{equation}
This form for the power spectrum illustrates several ideas in the
scattering theory approach. The partial wave index \(\ell\) is
directly related to the anomalous scaling exponent in the leading
power-law behaviour, and the momentum scale for the anomalous part of
the scaling relation is set by \(r_0^{-1}\). These observations
completely specify the meaning of the parameters \(\ell\) and \(r_0\)
in this formalism.  All deviations from scaling are contained in the
shape function, \(|F_\ell(k)/F_\ell(0)|\).

The usual form of the power-spectrum given in the literature is
non-dimensional, being a power-spectrum for a non-dimensionalized
curvature scalar \(\mathcal{R}\) \cite{Lidseyetal}.  Relating the
definition of \(\mathcal{R}\) to the objects given here, we find
\[
  P_{\mathcal{R}}(k) = 2\pi G\, P(k).
\]
This yields an object with no engineering dimensions. Note that we
have set \(\hbar\) and \(c\) to unity everywhere, so that \(G\) has
units of length-squared.  At this point it is possible to interpret a
large class of power-spectra in a model-independent way.  Next we
illustrate this approach by example.


\begin{figure}
\includegraphics[width=3.1in]{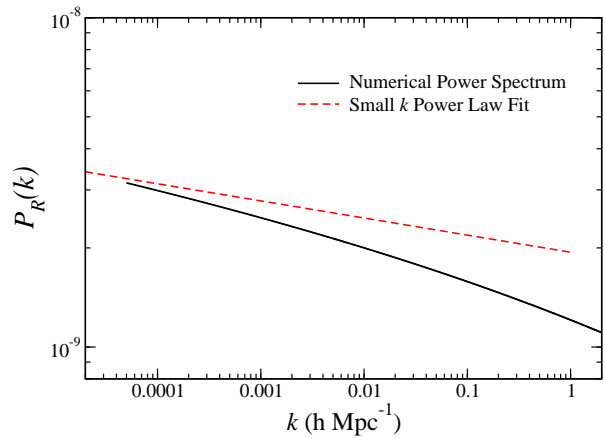}
\caption{Power spectrum computed for a \(\phi^2\) inflation model.
The dashed line indicates the small \(k\) behaviour, a pure power law.
The fitting procedure is discussed in the text.
}
\label{fig:powspec}
\end{figure}

As an example, consider the power spectrum in
Fig. \ref{fig:powspec}. This power spectrum was computed numerically
from an inflation model with a \(\phi^2\) inflaton
potential. Computations of power spectra for other single-field models
show that the properties of this power-spectrum are generic for a
large class of such models and it serves as a good illustration. The
behaviour as \(k\) approaches zero allows us to directly determine the
parameters \(\ell\) and \(r_0\).  The global behaviour is fit to the
\(k\rightarrow 0\) asymptotic form given by the combination of
Eqns.~(\ref{eqn:KlausEXT}) and (\ref{eqn:Pk}).  For fitting, one must
specify a range of \(k\) values in which the asymptotic result is
assumed to hold.  Using the full range of data for the power spectrum,
which is displayed in Fig. \ref{fig:powspec}, we obtain \(\ell \simeq
1.02496\), \(\delta \simeq 0.103\), and \(r_1 \simeq 6.43\times
10^{-15}\, h^{-1} \mathrm{Mpc}\,\mathrm{sec} / \mathrm{km}\).  Using a
significantly more restricted set of the data, with \(k < 0.0001\, h\,
\mathrm{Mpc}^{-1}\), we obtain \(\ell \simeq 1.02603\), \(\delta
\simeq 0.118\), and \(r_1 \simeq 1.39\times 10^{-13}\, h^{-1}
\mathrm{Mpc}\,\mathrm{sec} / \mathrm{km}\).  The large difference in
the determinations for \(r_1\) reflects the fact that the exponent
\(\delta\) is small; the actual coefficients of the corresponding
potentials differ only by about 10\%.

\begin{figure}
\includegraphics[width=3.1in]{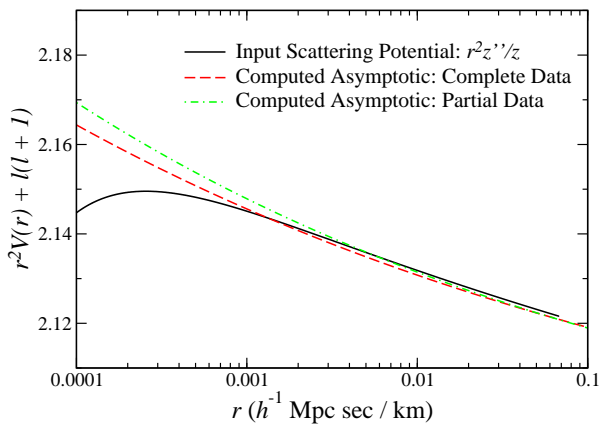}
\caption{Comparison of reconstructed scattering potential
to the input. The input was tabulated during computation of
the power spectrum.}
\label{fig:compare}
\end{figure}

With this information we reconstruct the large \(r\) behaviour for
\(V(r)\) and therefore \(z(r)\).  In Fig. \ref{fig:compare} we compare
the reconstructed function \(r^2 z''(r)/z(r) = r^2 V(r) +
\ell(\ell+1)\) to the input which gave the power spectrum of
Fig. \ref{fig:powspec}.  Note that both reconstructions agree well
with the input and with each other for large \(r\), as they should. A
small additive error results from lack of precision in the
determination of \(\ell\), which determines the pedestal for each of
the functions. The input function rolls over for small \(r\), which
indicates the end of inflation; of course, this rollover is not
captured by the asymptotic reconstruction.  Note further that the
amplitude of the power spectrum has no effect on these results; the
amplitude simply determines the scale \(r_0\). In this case \(r_0\) is
roughly \(5\times 10^5\, G^{1/2}\). The fact that \(r_0\) is much
larger than the Planck length reflects the fine-tuning necessary to
obtain a viable inflationary model.

To close the argument, we relate the large \(r\) asymptotic for
\(V(r)\) to the early time behaviour of the Hubble expansion.  If we
assume a past asymptotic for the Hubble expansion in the form
\[
  H(t) = p\, t^{-1} \left[1 + \left(\frac{t}{t_0}\right)^b +
  \dots\right], \quad t\rightarrow 0, 
\]
then a short calculation shows that
\[
  V(r) = \frac{1}{r^2} \left(\frac{r_1}{r}\right)^{b/(p-1)} + \dots,
  \quad r\rightarrow\infty, 
\]
where \(r_1\) can be written in terms of \(t_0\), though the relation
is not needed here. Applying this result to the example at hand, we
obtain \(b \simeq 4.2 - 4.5\). The large power is consistent with the
idea that the evolution of the scale factor for our \(\phi^2\) model
is very close to power-law inflation for most of the inflationary
epoch; this is why the power spectrum is close to a simple scaling
law.


Thus far we have discussed the power-spectrum for scalar fluctuations.
We now briefly show how to relate the scalar and tensor spectral
indices using the behaviour of the equation of state.

The stress-energy driving inflation can be parameterized by an
equation of state function \(w(\eta)\), defined by \(p = w \rho\).
Recall that one of the important predictions of inflation is a
relation between the spectral indices for scalar and tensor
fluctuation modes. During inflation the equation of state is close to
\(w = -1\). As inflation gradually ends, the equation of state drifts
away from \(-1\).  Assume a behaviour of the form
\[
  w(\eta) = -1 + (\eta_0/\eta)^{m} + \dots \quad, \eta\rightarrow
  -\infty. 
\]
We consider the simplest case for \(z(\eta)\), equivalent
to assuming that inflation is driven by a scalar-field.
In this case it is easy to show that the following relation holds:
\[
  \frac{a'}{a} = \frac{z'}{z} - \frac{1}{2} \frac{w'}{1+w}.
\]
Using this relation, a short calculation allows us to identify the
partial wave index for tensor perturbations as \(\ell_T = \ell -
m/2\), where \(\ell\) is the value for scalar perturbations. Therefore
we obtain a statement of the scalar-tensor relation in terms of the
parameter \(m\),
\[
  n_T = n_S - 1 + m.
\]
These spectral indices are defined as outlined in the above, in terms
of the small \(k\) behaviour of the power spectra, where the Jost
function contributions are negligible. For power-law inflation,
$\ell=\ell_T$, and we obtain the exact result $n_T=n_S-1$.


The procedure we have described is essentially non-parametric. No
particular model for the stress-energy need be assumed, and geometric
information is extracted directly from the power spectrum.
This analysis does not depend on slow-roll parametrizations
or assumptions, as distinct from previous treatments \cite{Lidseyetal,early}.
Equation (\ref{eqn:Pk}) provides a very clear separation of the contributions
to the power spectrum, and this separation enables the procedure.
Furthermore, the relations of scattering theory provide a very
convenient general framework for posing the density-perturbation
problem in inflationary cosmology.

We have seen that inflationary reconstruction is essentially a problem
in inverse scattering theory.  Direct relations were used to invert
from the power spectrum to asymptotic quantities of interest, which
describe the background evolution through most of the inflationary
epoch.  Although the full machinery of the Gelfand-Levitan equation
\cite{ChadanSabatier} was not introduced here, we are currently
investigating its use for future analyses. For example, the
Gelfand-Levitan equation can be applied to the analysis of any
features which do not represent small deviations from scaling, such as
might occur in multi-field scenarios or other dynamically complicated
models.

\end{document}